\documentstyle[preprint,aps]{revtex}
\begin{document}

\begin{title} 
{ 
C-Axis Electrodynamics as Evidence for the Interlayer Theory of
High $T_c$ Superconductivity
}
\end{title}

\author{ Philip W. Anderson}

\address {Joseph Henry Laboratories of Physics\\
Princeton University, Princeton, NJ 08544}
\maketitle
\vskip1truein
In the interlayer theory of high temperature superconductivity
 the interlayer pair
tunneling (similar to the Josephson or Lawrence-Doniach) energy
is the motivation for superconductivity. This connection requires two
experimentally verifiable identities. First, the coherent normal
state conductance must be too small relative to the ``Josephson''
coupling energy, and second,
the Josephson coupling energy must be equal to the condensation
energy of the superconductor. The first condition is well satisfied in 
the only case which is relevant, $(LaSr)_2CuO_4$; but the second
condition has been questioned. It is 
satisfied for all dopings in
$(La-Sr)_2CuO_4$, and also in optimally doped $Hg(Ba)_2CuO_5$ which was measured
recently, but seems to be strongly violated in measurements on
single crystals of $Tl_2Ba_2CuO_6$.


\vfill\eject

The theory that ascribes the phenomenon of high $T_c$
superconductivity in the cuprates primarily to interlayer
coupling\cite{1} correlates
electromagnetic coupling along the $c$-axis
(perpendicular, that is, to the $CuO_2$ planes) 
with the condensation energy of the
superconductor. This correlation, which should be particularly
sharp for ``one-layer'' materials, was proposed and roughly
tested against data on $(La-Sr)_2CuO_4$ (``$2\,1\,4$'') in the original
paper,\cite{2} and the equations were refined in papers by van
der Marel et al\cite{3} and Leggett\cite{4}.
In these latter papers, the apparent failure of
the relation in $Tl_2Ba_2CuO_6$ ``($Tl\ 2\,2\,0\,1$'') is emphasized, and 
new, rather
unequivocal measurements of $\lambda_c$, the $c$-axis penetration
depth by Moler et al\cite{5}, confirm this contradiction.
However, as I show below, there is
quite good agreement in a growing number of other cases:
$2\,1\,4$ at several different doping levels,\cite{6} and,
very recently, $Hg$ ``$1\,2\,0\,1$'' cuprate,
$HgCa_2CuO_4$.\cite{7} It appears then, that the $Tl$ salt is the
``odd man out'' or perhaps not a true one-layer case;
 this compound exhibits wide swings
in $T_c$ with preparation treatment.
Because both the $Tl$ and $Hg$ salts have relatively large $c$-axis
spacings and comparable $T_c$'s of around 90 K, the
contradiction between the two is particularly striking, and it is
only less important to confirm the measurements of Ref.\ [7],
preferably by another experimental method. 

Additional evidence for a major role for interlayer coupling is the
observation of a strong bilayer correlation in neutron scattering
in YBCO both in the superconducting state (for optimal
doping)\cite{8} and in the spin-gap regime,\cite{9} which is not
explicable in ``one-layer'' theories but receives a natural
explanation in the interlayer theory\cite{10}. Thus $Tl$
$2\,2\,0\,1$ stands out in providing contravening 
evidence against the theory of [1]. 

The ILT theory is very simple in principle. 
Electron motion for the cuprates in the $c$-direction is
incoherent in the normal state. 
This is unlike most normal metals, which are Fermi liquids and
which exhibit coherent transport in all directions. 
The interlayer hypothesis is that
electron pairing 
in the superconducting state makes this transport coherent, which
is actually observed, and which is responsible for the 
Josephson-like or Lawrence-Doniach-like superconducting coupling 
between the layers. In conventional superconductors, 
the Lawrence-Doniach coupling $\rm\underline {replaces }$ 
coherent transport in the
normal state, so that the superconductor gains no relative
energy, but in the cuprates, actual experimental observations 
exclude coherent transport in the normal state, so that the
$c$-axis energy is available as a pairing mechanism. [In my 
theory\cite{1}, the mechanism for blocking coherent transport is
the non-Fermi liquid nature of the normal metal state.] Thus
superconductivity occurs in connection with a two-dimensional (2D) to 3D
crossover; if for some reason, one desires a ``quantum critical
point'' to be associated with high $T_c$, that is its nature.

Our concept is, then, that there are two independent ways of
measuring the energy coupling the planes together in the
superconductor, each direct. The first is, in analogy with the
Josephson energy-current relation, to measure the electromagnetic
response to vector potentials along the $c$ axis, either by 
measuring the $c$-axis penetration depth or the $c$-axis
transverse plasma frequency. Because of the
correlation referred to above, this plasmon in the cuprates
lies, unusually, within the superconducting gap,  and is
clearly visible as a sharp edge in the reflectivity, followed by
a dip.

The second measurement is of the condensation energy of the
superconductor. Our postulate is that this is wholly, or almost
wholly, due to the $c$-axis coupling, so that it should be equal
numerically to the maximum possible value of electromagnetic
coupling, 
when all layers are equivalently coupled---that is,
only in ``one-layer'' superconductors. This is equivalent to the
statement that $\xi_c\simeq {c\over 2}$ where $\xi_c$ is the
c-axis correlation length. As pointed out in [2], 
this is a maximum possible value for $\xi_c$ or a minimum
for $\lambda_c$ for multilayer systems such as $YBCO$ or
$Bi_2Sr_2CaCu_2O_8$ (``BISCO $2\,2\,1\,2$'').  
The condensation energy may be estimated from
$T_c$ or $\Delta$, using BCS expressions, but because I am 
arguing that BCS does not use the correct form of interaction, 
it can give no better than an 
order of magnitude estimate, and
it is far better to use specific heat data when available. In
particular, the dependence on doping, according to such data, of
condensation energy is steeper than that of $T_c^2$ so that for
underdoped materials one  must be especially careful. $T_c$ is
roughly proportional to $x$, while $E_{\rm cond}$ is $\propto x^p$
with $p$ previously estimated as $\sim$ 3 to 4. 
The theory of Lee and Wen\cite{11} gives $p=3$ but has been
seriously questioned\cite{12}; specific heat data are therefore
more convincing, if hard to interpret quantitatively\cite{13}.
The sharp dependence of $\lambda_{ab}$ on $x$ which is predicted
by Lee and Wen in [11] $\rm\underline {is }$ 
observed in $2\ 1\ 4$\cite{13}
contrary to the criticisms of [12] which uses figures from
YBCO where effects of doping are less straightforward.
It is clear from such data that condensation energy falls off
with doping percentage $x$ more rapidly than $T_c^2$.

The basic formulas are as follows: First, for the electromagnetic
theory of an interlayer superconductor,  the basic London equation
is:
\begin{equation}
\vec {j}={1\over \lambda^2} \ \ {c\over 4\pi}\ \ \vec {A}
\end{equation}
where $c$ is the speed of light and $\vec {A}$ is the vector
potential. 
This is the definition of the penetration depth $\lambda$.
Focusing on $T<<T_c$, and ignoring the difference between free
energy F and energy E:
\begin{equation}
\vec {j}=c\, {\partial F\over \partial A}\simeq c\,{\partial E\over\partial
A}\end{equation}

The pairing energy in the interlayer theory comes entirely from
the coupling between planes, so that one can take $E$ to be the
condensation energy $E_b$ and assume the coupling energy has the
Josephson form 
\begin{equation}E_b=-E^\circ_b\ \cos\theta
\end{equation}
where $\theta$ is the phase difference between the pairs of
planes. In the presence of a vector potential
\begin{eqnarray}
\nabla\theta=&{2e\over \hbar c}A\\\nonumber
\theta=&{2ed\over\hbar c}A
\end{eqnarray}
where $d$ is the spacing between layers, $e$ is the charge of an
electron, and $\hbar$ is Planck's constant divided by $2\pi$. 
Combining Eqs.\ [2], [3] and [4]:

\begin{eqnarray}
j&=4c\,E^\circ_b \ \ \big ({e^2\,d^2\over \hbar^2\,c^2}\big)\, A \\\nonumber
\lambda_c&={\hbar c\over 2ed}\ \ {1\over \sqrt{4\pi
E^\circ_b}}\ \ .
\end{eqnarray}
A nearly equivalent measure of the electromagnetic coupling is
the $c$-axis plasma frequency. The dielectric constant $\epsilon$
is given
in terms of the $\delta$-function ``Drude weight'' 
\begin{equation}
\omega^2_p={c^2\over\lambda^2} 
\end{equation}
By 
\begin{equation}
\epsilon=-{\omega^2_p\over \omega^2}+\epsilon^0 
\end{equation}
and the edge occurs where $\epsilon$ changes sign, at 
\begin{eqnarray}
\hbar\omega^c_p&={\hbar
c\over\sqrt{\epsilon_0}\lambda}\\\nonumber
&=\sqrt{{4\pi E^\circ_b\over \epsilon_0}}\times 2ed
\end{eqnarray}
In the cases of $2\,1\,4$, $\sqrt {\epsilon_0}$ is an actually measured
quantity from the normal state reflectivity, because no appreciable
Drude weight appears in the normal state, and is $\sim 5\pm 1$.
In the cases of $Tl$ and $Hg$ one-layers, $\epsilon_0$ is not
well measured. $\lambda$, however, has no dependence on
$\epsilon_0$ and is the measured quantity in both of these 
cases. Figure 1  shows the
measured plasma edge for a series of doping levels in $2\,1\,4$. 

The thermodynamics of
optimally doped YBCO has been thoroughly studied by
Loram et al\cite{14} and their estimate for the condensation energy per
unit volume is:
\begin{equation}
E^\circ_b\,(YBCO)=3.5 \times 10^6\, {\rm erg\over cc}\ \ 
\end{equation}
(Per unit cell per layer, this is about 3 K, which is not
far from the BCS estimate of ${N(O)\,(kT_c)^2\over 2}$, taking
$N(O)$ to be $\sim$ 2 to 3/eV.) 

The binding energy for $2\,1\,4$ must be estimated from Loram et
al's curves\cite{13,14} (Figure 2),  which also,
fortunately, shows several doping levels. 
For the optimal doping level, 17 to 20\%,  
$E_b$ can be estimated with the identities: 
\begin{equation}
\int (c_N-c_s)\, dT = E_b =\int T\Delta\gamma\,(T)\,dT
\end{equation}
\begin{equation}
\int (c_N-c_s){dT\over T}=0=\int\Delta\gamma\,(T)\,dT
\end{equation}
where $\gamma$ is the quantity $c/T$ plotted in Figure 2, 
and $c_N$ is the normal and $c_s$ the superconducting specific
heat. 
The total binding energy is considerably smaller than that of
YBCO, roughly
\begin{eqnarray}
E^0_b\simeq &220\pm 50 {{\rm mJ}\over {\rm gm\ atom}}\\\nonumber
=&1.7\pm .4\times 10^5 {\rm erg/cc}
\end{eqnarray}
Interestingly, this is below (by a factor of 2) what I would
predict from scaling from YBCO by $T_c^2$, 
perhaps partly because of a contribution from the chains. 
With less accuracy because of the critical fluctuation effects
on $c_s$ 
for low doping, I can also estimate $E_b$ for the doping
levels 13.5\% and 10\%. 

Using the value (13) I obtain $\lambda_c=3\pm 1\mu m$ for optimal 
doping, which is embarrassingly close. 
Figure 3 shows my estimates
for three doping levels plotted on Uchida's curve of $\lambda$ as
calculated from Figure 1 using Eq.\ 9, with my estimates
plotted as areas that make some attempt to express the
uncertainties.

The agreement both as to numerical value and trend
is heartening. For $2\,1\,4$, driving a critical
Josephson current is precisely sufficient to erase the energy of
the superconducting correlation. Undoubtedly it is possible to
invent a system of carefully balanced cancellations that would
nonetheless ascribe the source of superconductivity to internal
correlations in the planes but such logical contortions seem
improbable and may even be impossible.
Why would an intraplanar mechanism 
correlate its $T_c$ as well
as its energy precisely with the strength of 
interplanar
coupling, over a range of 5 to 1 in $T_c$?

The case of $Hg$ $1\,2\,0\,1$ is much less airtight, but still
strong. I know of no satisfactory specific heat measurements so we
are reduced to scaling the binding energy according to $T^2_c$, 
and hence $\lambda_c$ according to $T_c$. I predict, then,
for $Hg$ $1\,2\,0\,1$
\begin{eqnarray}
\lambda_c&=3\mu m\times{40\over 90}\times{d_{214}\over
d_{Hg}}\\\nonumber
 &=1\pm 0.5\mu m
\end{eqnarray}
The observed value\cite{7} is quoted as $1.34\mu \pm $ about
10\%. The agreement is spectacular. 

$Tl$ $1\,2\,0\,1$ would be predicted, on the same basis, to have
$\lambda_c\simeq 0.8\mu m$, since $d$ is even greater than that for
$Hg$, 
but K. Moler et al\cite{5}
find that $\lambda_c> 15\mu m$ for the single crystals for which they
have imaged vortices, and this figure is in agreement with
estimates by van der Marel (and with my own estimates
using transport theory). This is clearly a severe anomaly. The above
direct evidence for interplanar coupling in the other cases is
supplemented by the neutron scattering evidence in YBCO which
shows that the gap structure is strongly correlated between
planes in the close pair, in just such a way as to optimize
interplanar kinetic energy\cite{10}. I cannot emphasize too strongly the
need to assure ourselves that $Tl$ $1\,2\,0\,1$ is genuinely a
one-layer case. Some evidence for structural defects exists. 

\vskip.3truein


\medskip
The ILT hypothesis for the high $T_c$ cuprates was based
from the start on an experimental observation: that $c$-axis
conductivity is non-metallic and incoherent where that in the
$ab$ plane is metallic, if in many respects very anomalous. This
behavior is presumed to be a result of a non-Fermi liquid,
charge-spin separated state; but the hypothesis can be directly
tested in a manner completely independent of that conjecture.
There are two experimentally testable consequences of the idea,
if one is able to measure the $c$-axis electrodynamics in the
superconducting state, as has been done in a number of cases. The
first is violation of the ``Josephson identity'', which
expresses the fact that in BCS superconductors pair tunneling
replaces the coherent normal state conduction. This violation 
has been noted previously by Timusk.\cite{15}  The
second is the requirement that the supercurrent kernel 
${c\over 4\pi^2\lambda^2}$ almost precisely match the condensation energy of the
superconductors. It seems to
me that this agreement effectively rules out any intralayer
theory of high $T_c$, and points to the interlayer
concept, for those cases in which it occurs; but we are left at a
loss in the one clear case where it does not\cite{16}.

\vfill\eject

\centerline{\bf FIGURE CAPTIONS}
\bigskip

\begin{enumerate}

\item {Reflectivity measurements from ref.\ (6)}

\item{Specific heat of $2\,1\,4$ samples. Doping $x$ is the
      parameter. }

\item{The points are the measured values of $\lambda_c$ from
ref.\ (6); the large ovals are the result of our theory (eq.\
[6]) including rough estimates of limits of error. }

\end{enumerate}

\vfill\eject

\end{document}